\documentclass[aip,amsmath,amssymb,reprint,floatfix]{revtex4-1}

\usepackage[utf8]{inputenc}
\usepackage[T1]{fontenc}
\usepackage{mathptmx}
\usepackage{etoolbox}
\usepackage{graphicx}
\usepackage{dcolumn}
\usepackage{bm}
\usepackage[caption=false]{subfig}
\usepackage{lipsum}
\usepackage{amsmath,amssymb}
\usepackage{placeins}
\usepackage{balance}
\usepackage{dblfloatfix}

\makeatletter
\def\@email#1#2{
 \endgroup
 \patchcmd{\titleblock@produce}
  {\frontmatter@RRAPformat}
  {\frontmatter@RRAPformat{\produce@RRAP{*#1\href{mailto:#2}{#2}}}\frontmatter@RRAPformat}
  {}{}
}
\makeatother
\begin{document}

\title[Moderator Simulation for the Compact Positron Source at NLCTA]{Moderator Simulation for the Compact Positron Source at NLCTA}
\author{Ryland Goldman}
\email{rgoldman@slac.stanford.edu}
\affiliation{SLAC National Accelerator Laboratory, Menlo Park, California 94025, USA}
\affiliation{Department of Physics and Astronomy, University of California, Los Angeles, Los Angeles, California 90095, USA}

\author{Sophie Crisp}
\affiliation{SLAC National Accelerator Laboratory, Menlo Park, California 94025, USA}

\author{Spencer Gessner}
\affiliation{SLAC National Accelerator Laboratory, Menlo Park, California 94025, USA}

\date{\today}

\begin{abstract}
Positron moderation is a key part of the production of slow positrons. This paper examines G4beamline simulations of various moderator designs and the addition of a radiofrequency cavity to increase the efficiency of the moderator. A setup consisting of a 100~MeV electron beam incident on a 7~mm tungsten target, a 5~T adiabatic matching device, a 5-cell L-band RF cavity, and a single-crystal tungsten moderator leads to a 39.3$\times$ improvement in the number of positrons moderated over a common thin-foil polycrystalline moderator without the AMD and cavity. We also present a novel calculation method for determining the stopping distribution of low-energy positrons in a thin foil, modifying the Makhovian distribution with an exponential ramp to account for edge effects.
\end{abstract}

\maketitle

\section{Introduction}
Slow positrons are positrons with kinetic energy $<30$ keV. Among other fields, these positrons have applications in positron annihilation spectroscopy, ultrafast positron diffraction, and antimatter research. The Next Linear Collider Test Accelerator (NLCTA) at SLAC is currently in the planning stage for a compact positron source that produces slow positrons from an electron linac.

A target (also referred to as a ``converter'' by some sources) produces an electromagnetic shower when the high-energy electrons interact with high-$Z$ nuclei through Coulomb scattering. The resulting Bremsstrahlung radiation is capable of producing electron-positron pairs. These positrons are collimated by an adiabatic matching device and then decelerated using a radiofrequency cavity.

We then capture these positrons by stopping them inside a tungsten moderator, where the negative work function of the material causes some to be reemitted from the surface with 3.0 eV kinetic energy.~\cite{HUGENSCHMIDT2002283} This project attempts to optimize the positron yield from a target-moderator system.

\begin{figure}[htbp!]
    \centering
    \includegraphics[width=\linewidth]{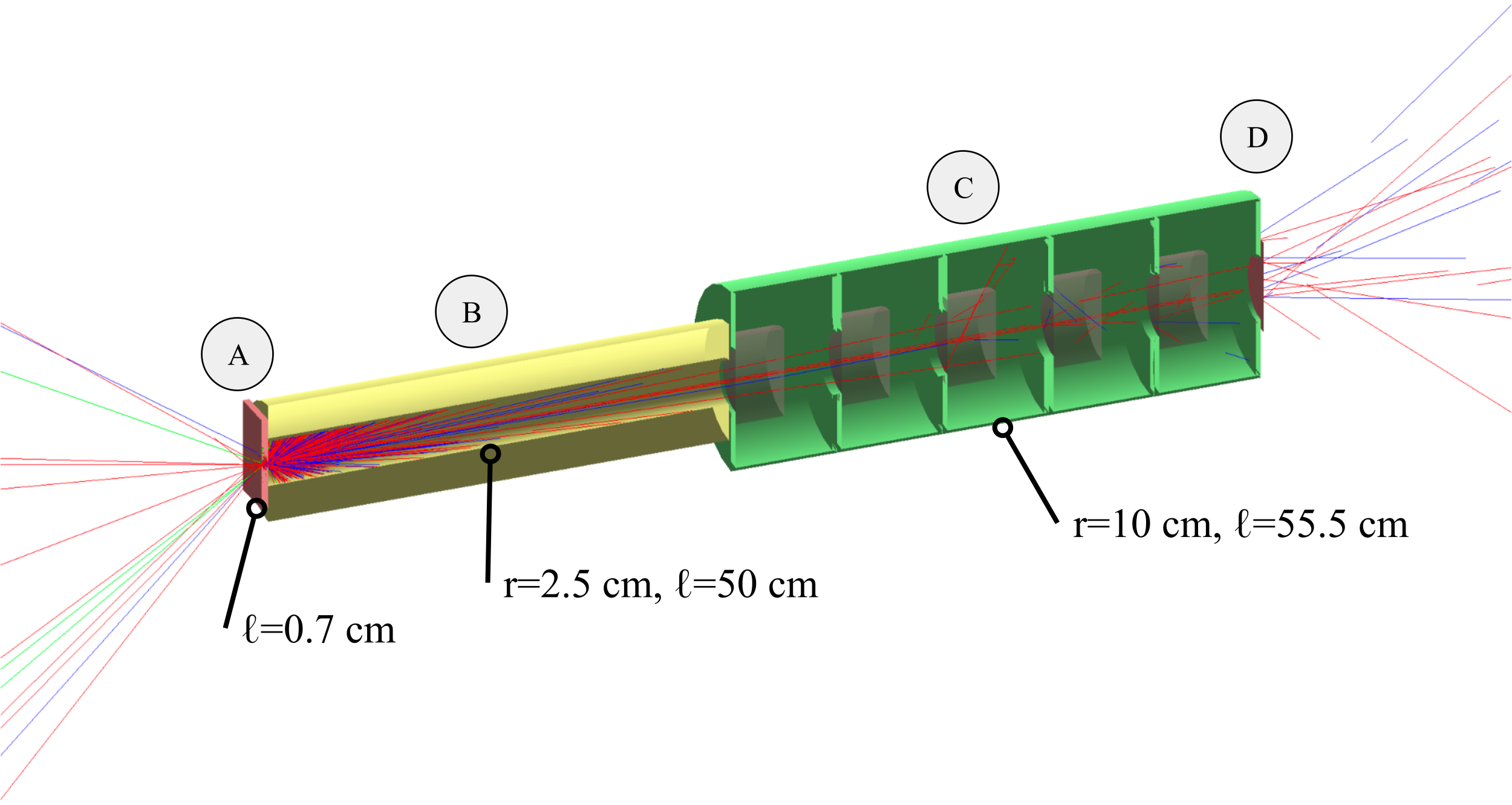}
    \caption{Schematic of the full linac setup, including the target (A), adiabatic matching device (B), RF cavity (C), and moderator (D). The particle trajectories are colored red for electrons, blue for positrons, and green for photons (photons emitted past the target are excluded for visibility).}
    \label{fig:schematic}
\end{figure}

A schematic of the setup is shown in Fig.~\ref{fig:schematic}.

\section{Methods}

The simulations from this project use G4beamline, a particle tracking program derived from the GEANT4 toolkit.~\cite{G4beamline,GEANT4} The Penelope2008 models~\cite{Salvat2009} are used via the FTFP\_BERT\_PEN physics package. Most simulations were parallelized to run on a 192-core c6a.metal instance on AWS EC2.

The key metrics include the target positron yield $\gamma_t$, which represents the fraction of electrons converted to positrons, the moderator yield $\gamma_m$, which represents the number of positrons moderated as a fraction of incident positrons, and the overall system yield $\gamma$, which is the ratio of the total number of moderated positrons to input electrons.

\section{Target}
\begin{figure}[htbp!]
    \centering
    \includegraphics[width=\linewidth]{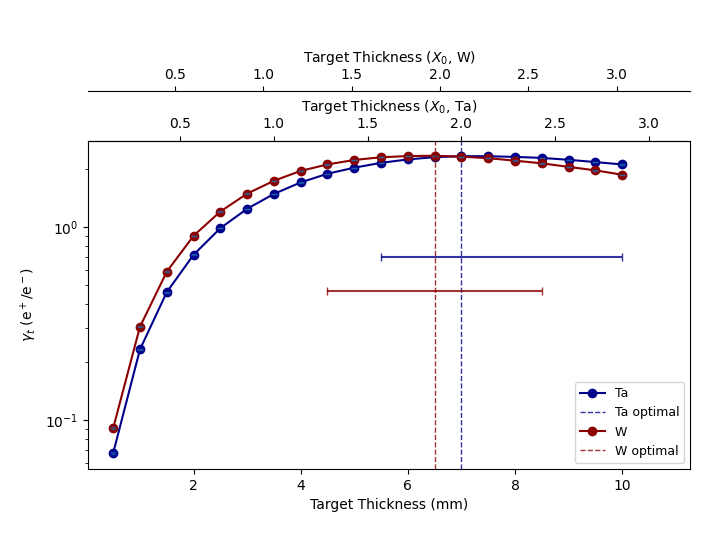}
    \caption{Positron yield $\gamma_t$ as a function of target thickness (shown in both millimeters and radiation lengths) for tungsten and tantalum. The vertical lines represent the optimal thickness and error bars show thicknesses within 90\% of the optimal yield. For tungsten, the optimal thickness is between 4.5 and 8.5~mm. For tantalum, it is between 5.5 and 10.0~mm. We use 100 MeV electrons; see~\cite{ORourke}~for a comparable plot with different electron energies.}
    \label{fig:yield_vs_thickness}
\end{figure}

Initially, the target parameters (thickness and composition) are selected to optimize the positron yield $\gamma_t$, defined as the fraction of positrons emitted from the target relative to the number of electrons in. We simulate a 100 MeV electron beam directly incident on a target of varying thickness. The number of positrons produced in the forward direction is divided by the number of incident electrons as shown in Fig.~\ref{fig:yield_vs_thickness}, which agrees with the data found by previous groups.~\cite{ORourke}

The optimal thickness is found to be 6.5~mm for tungsten and 7.0~mm for tantalum, and the yield at optimal thickness to be $0.2330\pm0.0013$ positrons per incident electron for tungsten compared to $0.2318\pm 0.0013$ for tantalum. The range of thicknesses which deliver at least 90\% of the maximum yield is 4.5-8.5~mm for tungsten and 5.5.0-10.0~mm for tantalum.

As there is no significant performance difference between the materials, we instead consider the material properties. Because of its melting point, the GBAR project at CERN chose to use tungsten to prevent ablation by the electron beam.~\cite{GBAR, Liszkay} Tungsten is also significantly less expensive than tantalum.~\cite{Liszkay}

\begin{figure}[htbp!]
    \centering
    \includegraphics[width=\linewidth]{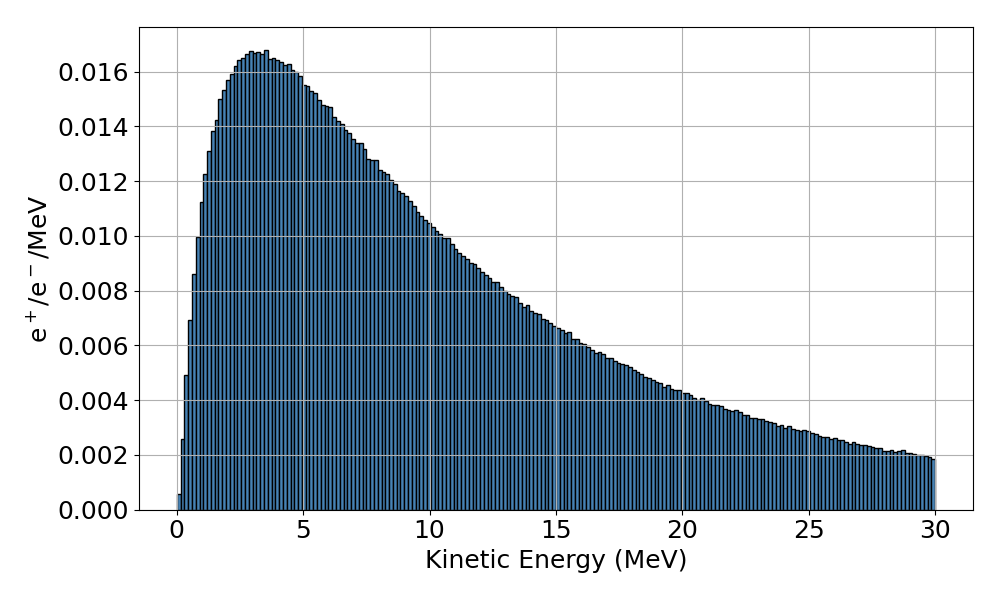}
    \caption{Energy distribution histogram of positrons emitted in the forward direction from 100 MeV electron collisions with a~7.0~mm tungsten target.}
    \label{fig:energytarget}
\end{figure}

Therefore, we select a tungsten target with a thickness of 7~mm, which is used for the remainder of this paper. Fig.~\ref{fig:energytarget} shows the energy spread of the resulting positrons, which peaks at approximately 3 MeV. This peak is dependent on the thickness of the target and energy of the electron beam.

\section{Adiabatic Matching Device}
The positrons emitted from the target have a wide angular divergence which must be reduced in order for the beam to pass through the RF cavity. We use an adiabatic matching device (AMD), which consists of a solenoidal magnetic field decreasing from 5.0~T to 0.5~T with the formula,
\begin{eqnarray}
    B_z(\rho=0)=\frac{B_0}{1+\alpha z}\\
    B_\rho(\rho,z)=\frac{B_0\alpha\rho}{2\left( 1+\alpha z \right)^2},
\end{eqnarray}
where $\alpha=0.018$~mm$^{-1}$ is the tapering parameter.~\cite{AMD,AMDslac}

\begin{figure}[htbp!]
    \centering
    \includegraphics[width=\linewidth]{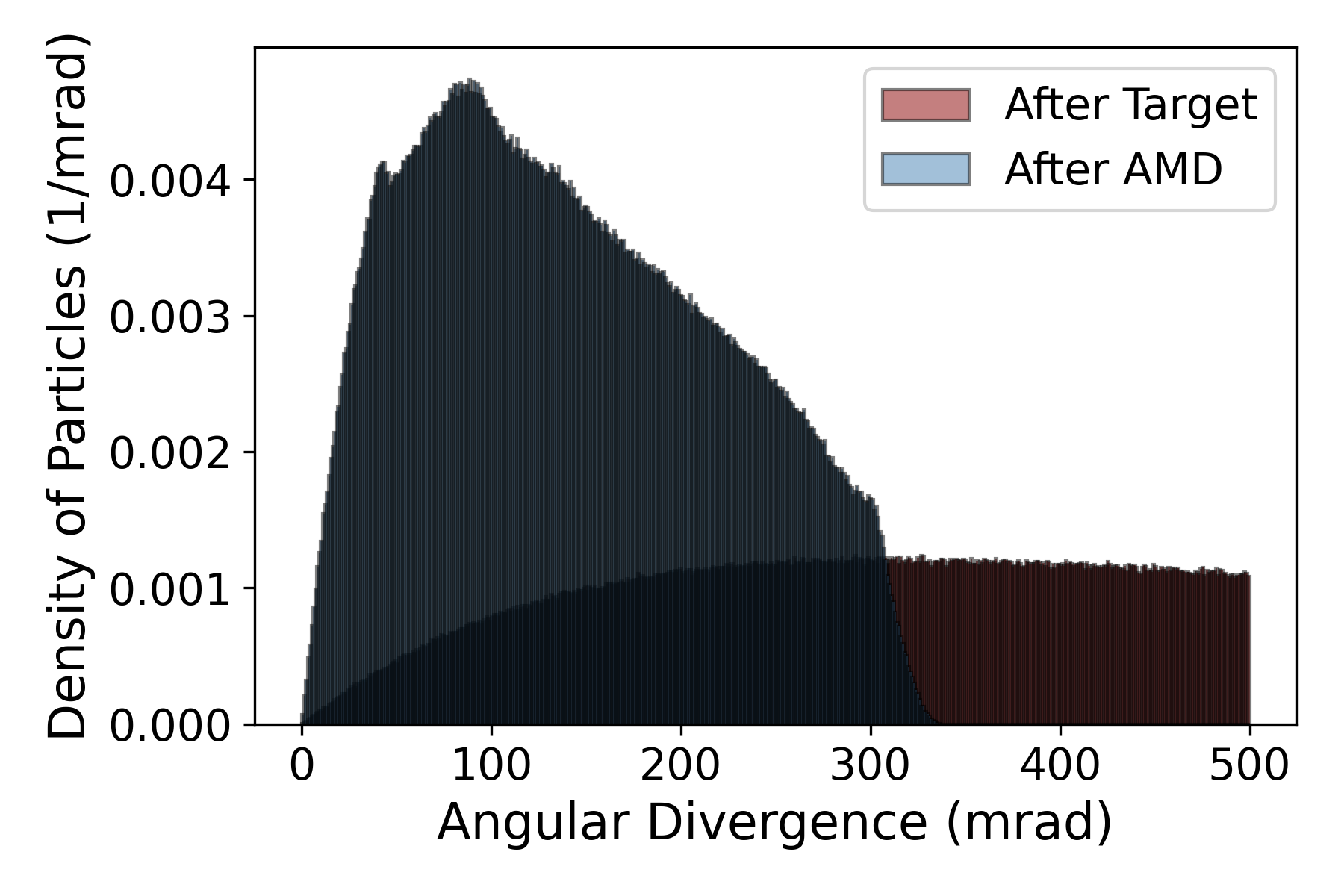}
    \caption{Angular divergence histogram of positrons before and after passing through the adiabatic matching device. The AMD reduces the spread of nearly all particles to under 300 mrad. The reduced angular divergence allows particles to pass through the RF cavity without hitting the walls.}
    \label{fig:AMD}
\end{figure}
\begin{figure}[htbp!]
    \centering
    \includegraphics[width=\linewidth]{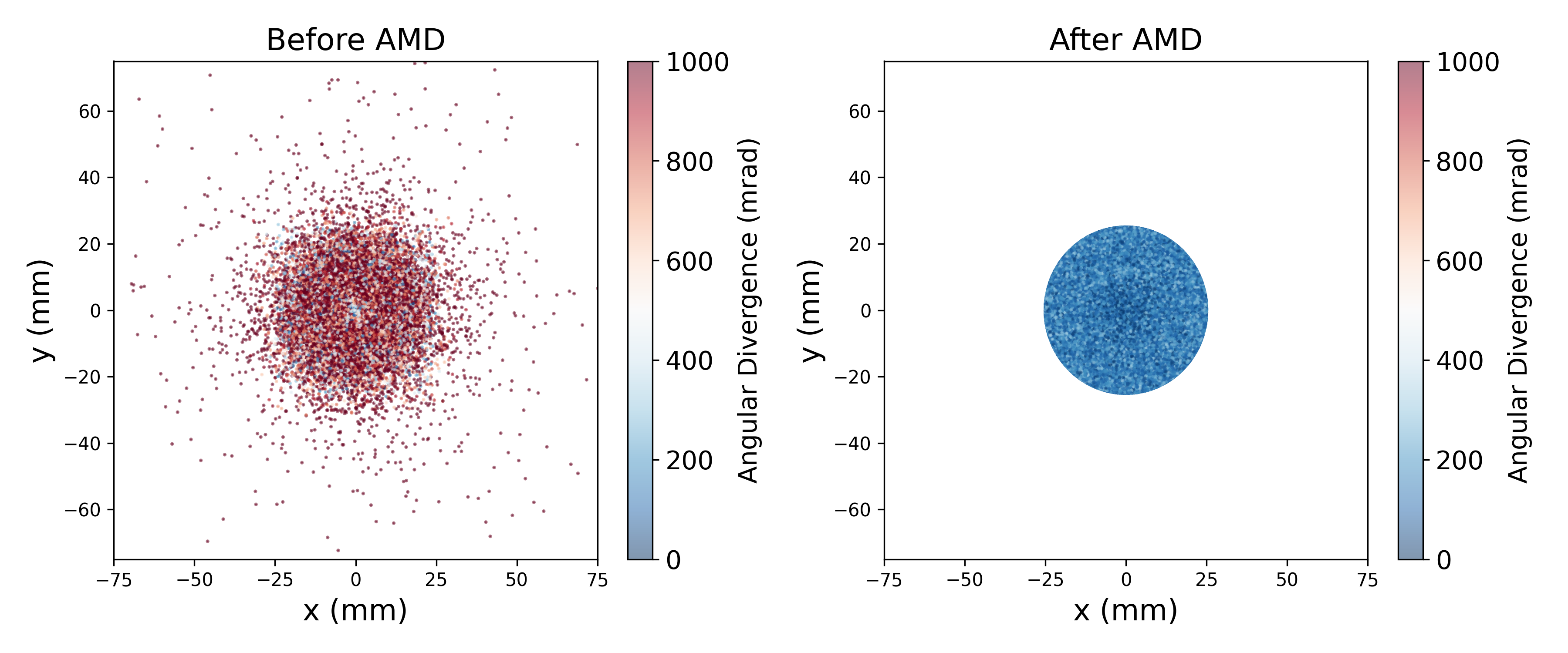}
    \caption{Cross-section of positron beam a) emitted by the target, and b) after leaving the AMD, colored by angular divergence. The AMD was modeled as a 25~mm cylinder; particles outside of this radius were killed.}
    \label{fig:AMD2}
\end{figure}

As shown in Fig.~\ref{fig:AMD}, the AMD reduces the beam's angular divergence so particles can enter the small aperture of the RF cavity. Fig.~\ref{fig:AMD2} additionally shows the beam's transverse displacement. Particles with $r>25$ mm are removed from the simulation as they are assumed to come in contact with the edge of the AMD.

Of the high-energy positrons emitted from the target, 69.9\% survive the AMD and pass into the RF cavity.

\section{Decelerating RF Cavity}
To decelerate the beam, we use a five-cell $L$-band (1.3 GHz) radiofrequency cavity previously developed for the ILC linac~\cite{PhysRevSTAB.12.042001}. The cavity has been tested up to a gradient of 13.7 MV/m, but we simulate it at a design gradient of 15 MV/m.

The search method used to locate the optimal timing offsets for each cell was a Monte Carlo simulated annealing algorithm. Several random initial conditions are chosen throughout parameter space. Similar to a gradient descent, neighboring points are evaluated for the cost function (for example, maximizing particles under 200 keV), but less optimal solutions are accepted probabilistically to escape any local minima. The step size and acceptance probability (temperature) are decreased over time to mimic the annealing of materials. This method is preferable to a gradient descent as it requires less compute time.

\begin{figure}[htbp!]
    \centering
    \includegraphics[width=\linewidth]{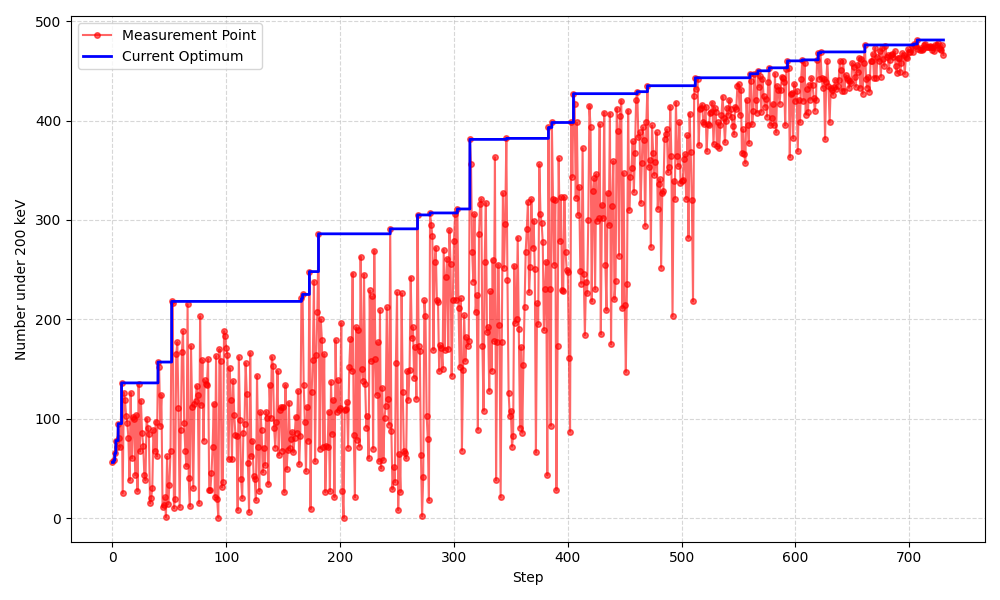}
    \caption{The simulated annealing algorithm used to tune the RF cavity. Here, the goal was to optimize the number of positrons under 200 keV, which are the most likely to stop in the moderator. Different goals were used to produce the greatest yield in each moderator.}
    \label{fig:SimulatedAnnealing}
\end{figure}

\begin{figure}[htbp!]
    \centering
    \includegraphics[width=\linewidth]{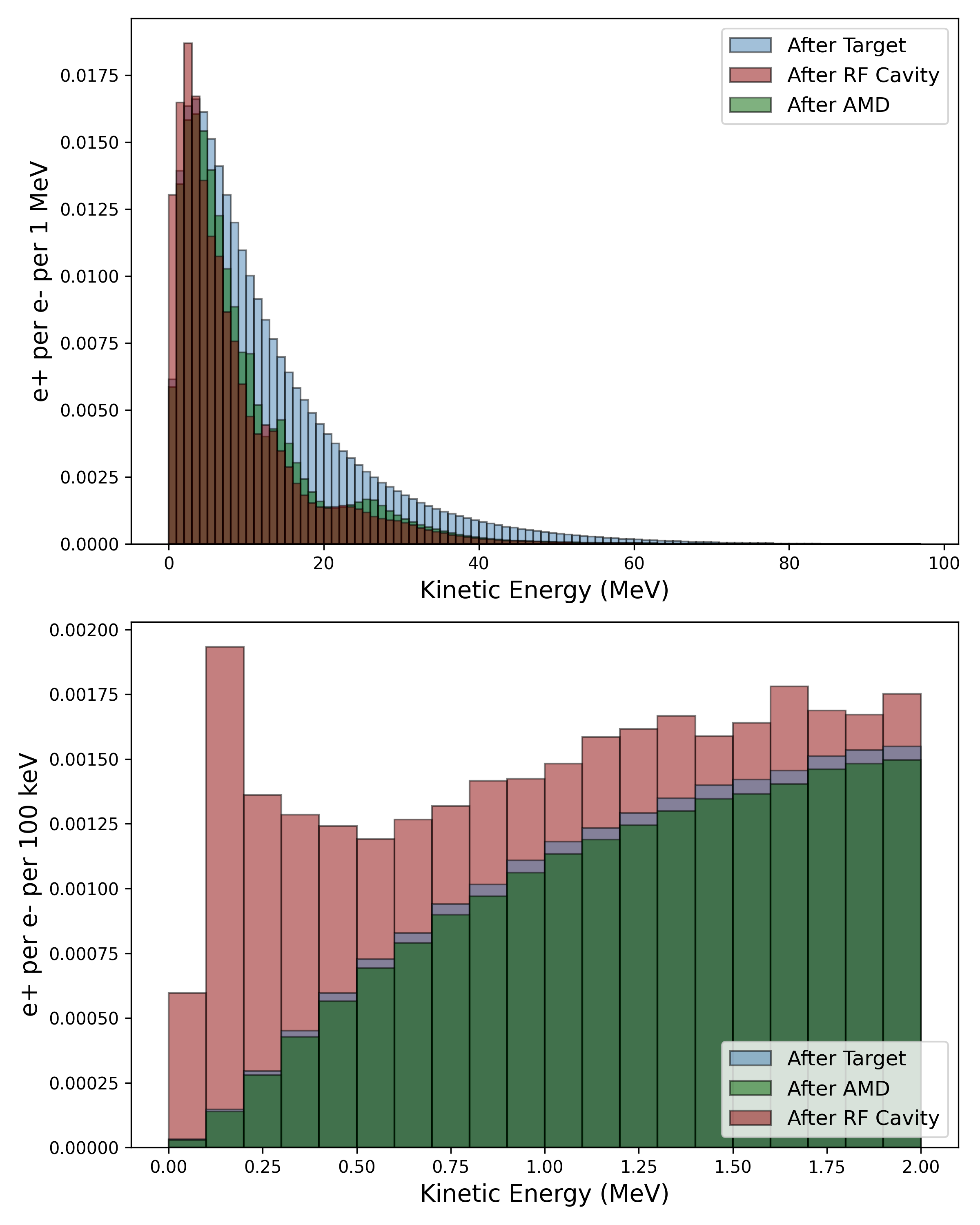}
    \caption{Resulting energy distribution of positrons before and after passing through the RF cavity. The distribution has been shifted towards lower energies.}
    \label{fig:LBand}
\end{figure}
\begin{figure}[htbp!]
    \centering
    \includegraphics[width=\linewidth]{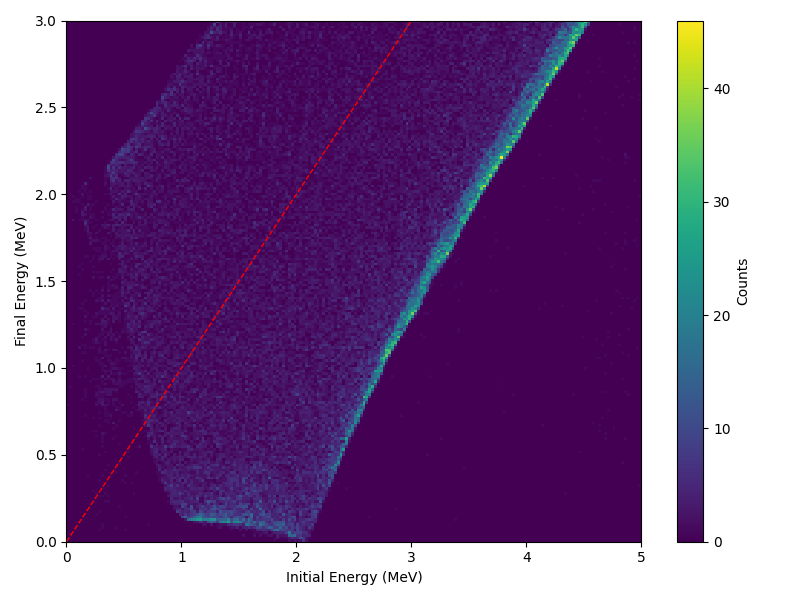}
    \caption{Energy change from particles before and after the RF cavity. The diagonal red line represents no change. Particles above the red line have increased energy, while those below the line have decreased energy.}
    \label{fig:EnergyChange}
\end{figure}

Figs.~\ref{fig:LBand} and~\ref{fig:EnergyChange} show the result of this optimization, with the end parameters listed in \ref{tab:cavityparam}. The cavity increased the number of positrons with $KE<200$~keV by a factor of 15.

\begin{table}
\centering
\caption{RF cavity simulation parameters. These are the optimal raw timing offsets provided to G4beamline and found using the simulated annealing algorithm. It is not guaranteed to be the global best solution, only a reasonably good solution. At 1.3 GHz, a period of $2\pi$ is 0.769 ns.}
\label{tab:cavityparam}
\begin{ruledtabular}
\begin{tabular}{cccc}
&Cavity & Phase (ns) \\
\hline
&1 & $0.732331$ & \\
&2 & $0.159283$ & \\
&3 & $0.172267$ &  \\
&4 & $0.361927$ & \\
&5 & $0.275202$ \\
\end{tabular}
\end{ruledtabular}
\end{table}

\begin{figure}[htbp!]
    \centering
    \includegraphics[width=\linewidth]{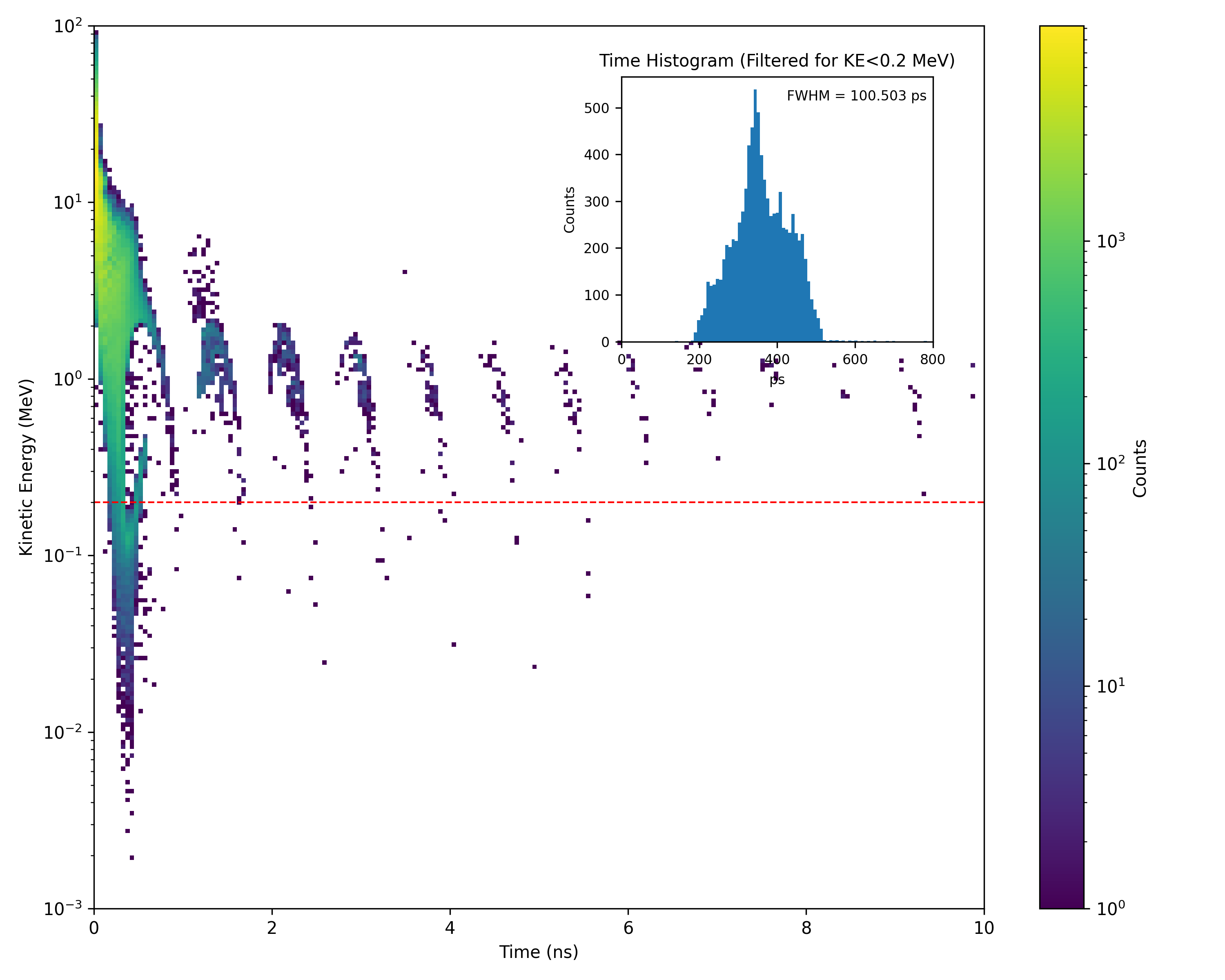}
    \caption{Temporal spread of particles after exiting the RF cavity, by energy. The inset histogram filters for positrons above 200 keV.}
    \label{fig:RFTemporalSpread}
\end{figure}

The temporal spread after the RF cavity is shown in Fig.~\ref{fig:RFTemporalSpread}.

\section{Moderator Theory}
In matter, positrons lose energy due to Bremsstrahlung radiation, Coulomb scattering, and ionization. Most low-energy particles (and a fraction of mid- to high-energy particles) will come to a stop inside of the material. Once stopped in a tungsten foil, they are subject to thermodynamic effects. Positrons which stop closest to the edge of the foil are more likely to be expelled through this thermalization process.

\subsection{Makhovian Distribution}
The stopping distribution of positrons in a thick material is described by the Makhovian distribution,
\begin{equation}
    p(z)=\frac{m\,z^{m-1}}{(z_0)^m}\exp\left( \left( -\frac{z}{z_0} \right)^m \right),
\end{equation}
where $p(z)$ is a probability density function and $m$ and $z_0$ are constants determined by the positron's angle of incidence and energy.~\cite{Alsulami}

\begin{figure}[htbp!]
    \centering
    \subfloat[Plot of $m$\label{fig:MakhovianParamsM}]
        {\includegraphics[width=0.48\linewidth]{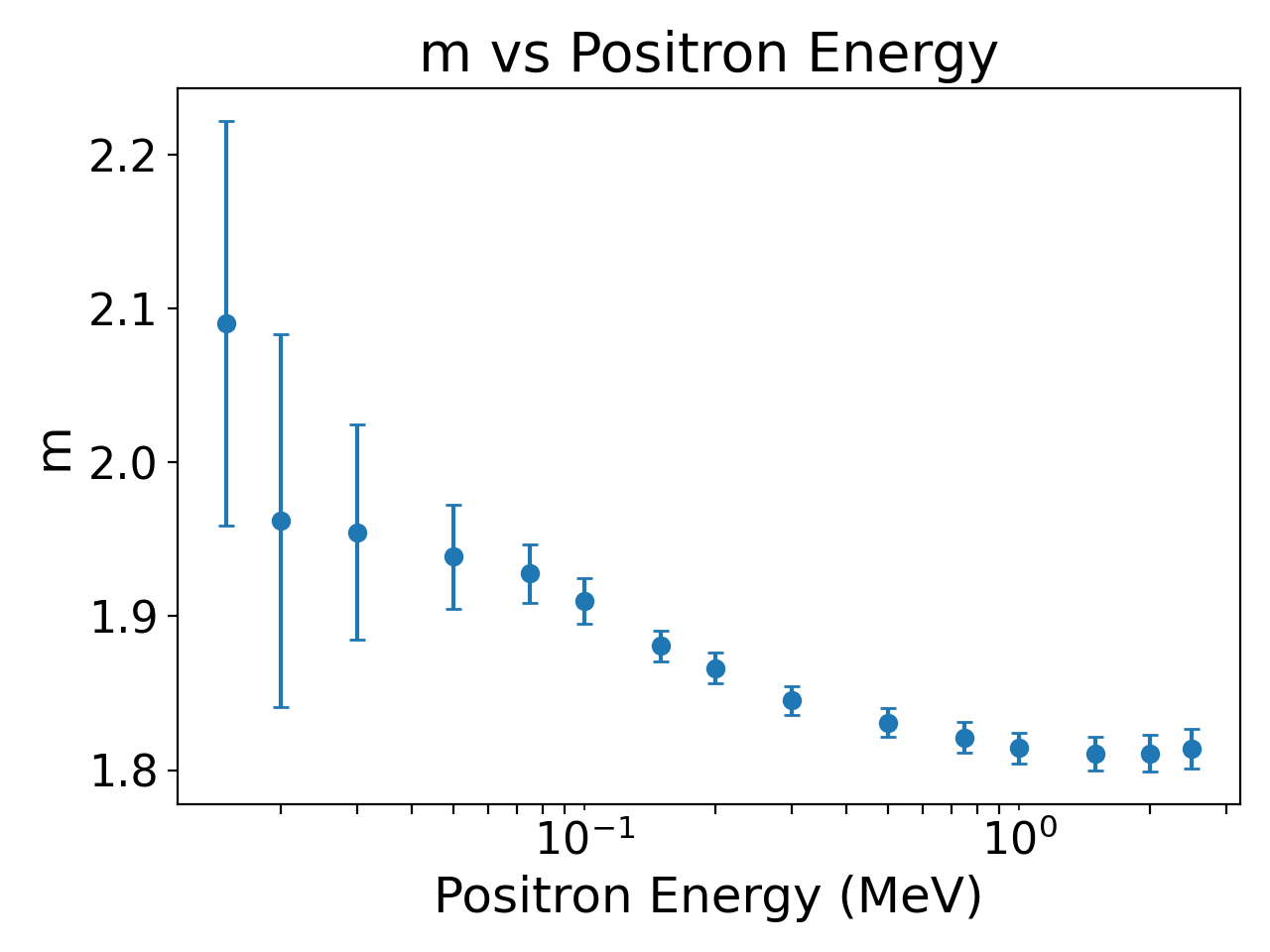}}\hfill
    \subfloat[Plot of $z_0$, with fit\label{fig:MakhovianParamsZ}]
        {\includegraphics[width=0.48\linewidth]{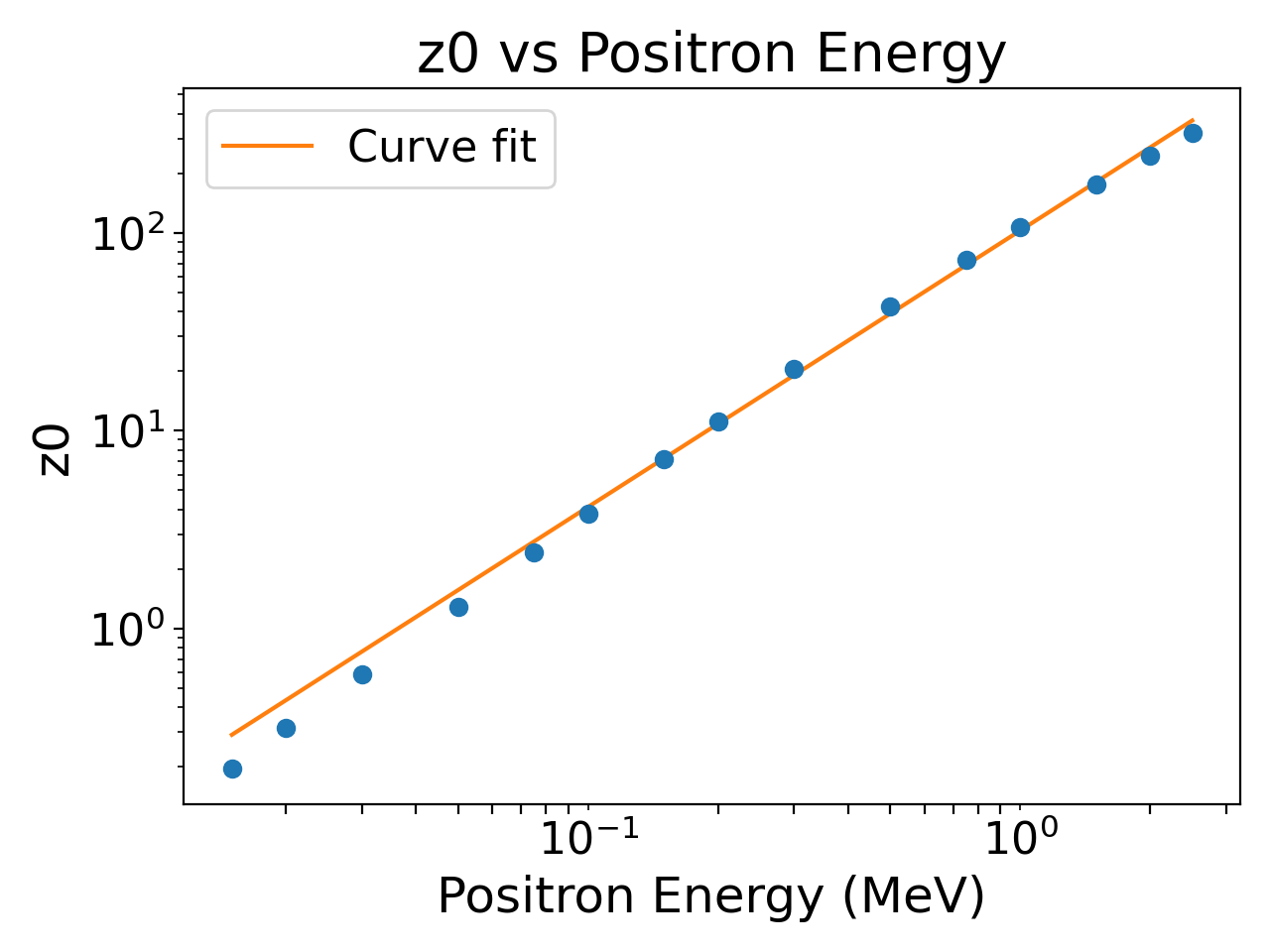}}
    \caption{Makhovian distribution parameters $m$ and $z_0$ as a function of initial positron kinetic energy. $z_0(E)$ has been fit to a power function $ax^n$, shown as the orange line.}
    \label{fig:MakhovianParams}
\end{figure}

For normal incidence, $m\approx1.8$, with the values for $z_0$ as listed in Table~\ref{tab:makhov}. $z_0$ approximately follows a power-law curve of the form $ax^n$, where $a=102.8\pm1.8$ and $n=1.397\pm0.004$, as shown in Fig.~\ref{fig:MakhovianParamsZ}.

\begin{table}
\centering
\caption{Makhovian distribution parameter $z_0$ for tungsten for varying positron kinetic energies (KE) at normal incidence. There is variation between physics packages and these constants may not perfectly match those from other simulation software packages.~\cite{DRYZEK20084000}}
\label{tab:makhov}
\begin{ruledtabular}
\begin{tabular}{cccc}
&KE (MeV) & $z_0$ (µm) \\
\hline
&0.1 & $3.79\pm0.021$ & \\
&0.5 & $42.46\pm0.16$ & \\
&1.0 & $106.46\pm0.44$ &  \\
&5.0 & $649.72\pm 1.13$ & \\
&10.0& $1349.36\pm1.78$ \\
\end{tabular}
\end{ruledtabular}
\end{table}

\subsection{Optimization}
\begin{figure}[htbp!]
    \centering
    \includegraphics[width=\linewidth]{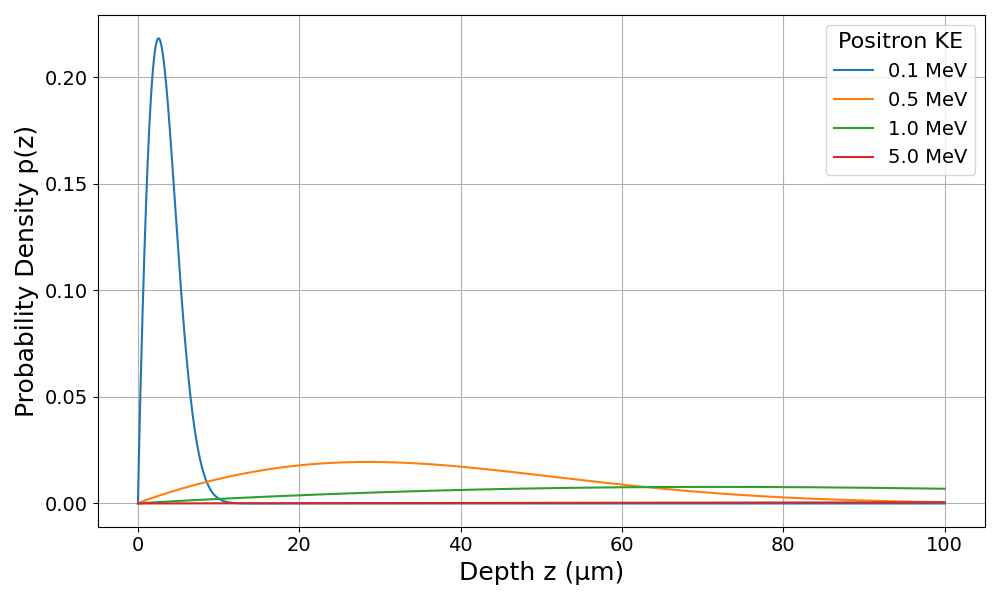}
    \caption{Makhovian stopping distributions at normal incidence for different positron energies. As expected, lower energy positrons stop much closer to the edge, which increases the probability of thermalization. Therefore, the RF cavity attempts to decrease the energy of the particles by targeting the greatest number under 200 keV.}
    \label{fig:Makhovian}
\end{figure}

\begin{figure}[htbp!]
    \centering
    \includegraphics[width=\linewidth]{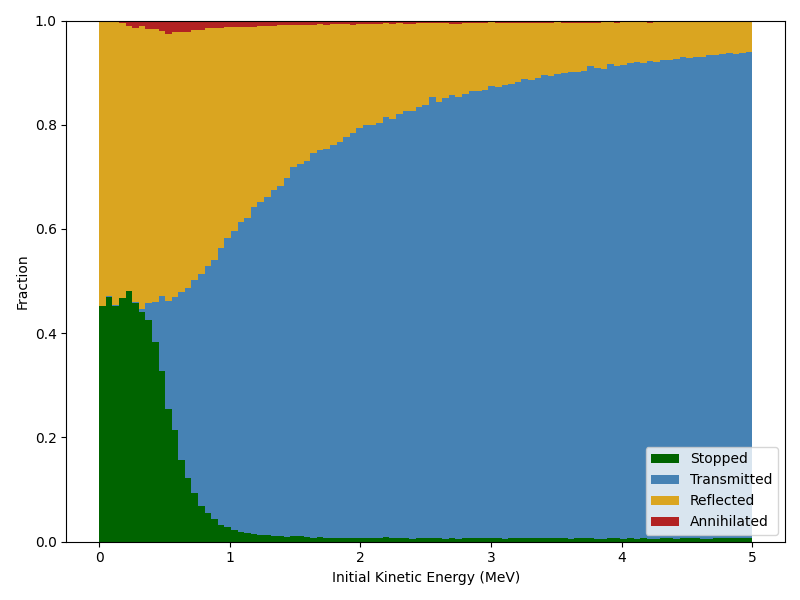}
    \caption{Outcome of incoming positrons in a 50 µm foil moderator, categorized by initial kinetic energy. Positrons with energy over 1 MeV essentially do not contribute to the moderator yield.}
    \label{fig:FoilEnergy}
\end{figure}

Sample values for the Makhovian distribution at different energies are shown in Fig.~\ref{fig:Makhovian}. For a beam with a wide energy spread, only the low-energy particles are stopped in the moderator (Fig.~\ref{fig:FoilEnergy}). 27.2\% of positrons with kinetic energies under 500 keV stop. Additionally, many of these low-energy positrons are reflected off of the moderator surface.

To ensure that the greatest quantity of positrons stop in the moderator, the energy of the beam must be reduced using a decelerating cavity (see Section VI).

\subsection{Ramped Makhovian}
\begin{figure}[htbp!]
    \centering
    \includegraphics[width=\linewidth]{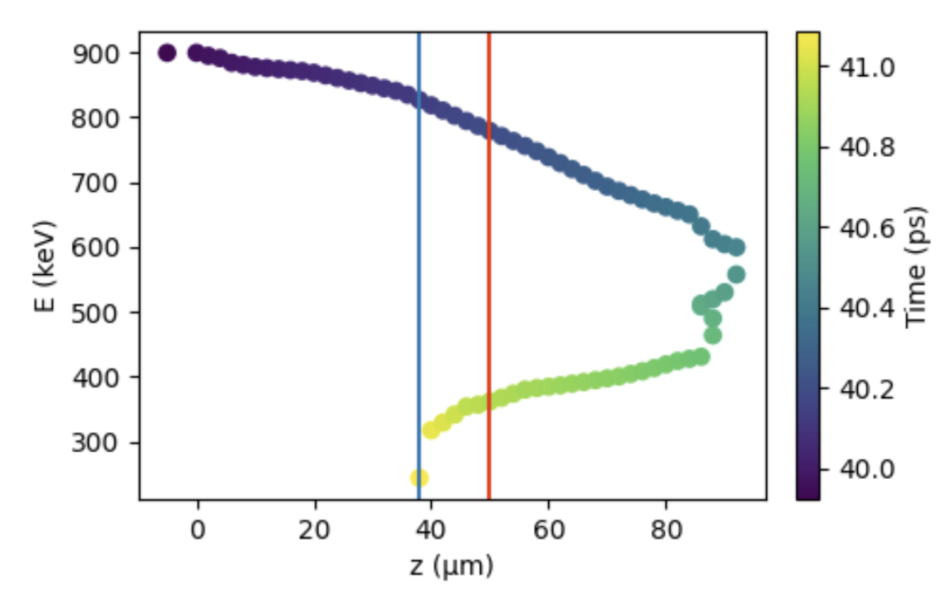}
    \caption{Particle trace of an incident positron that would contribute to the Makhovian, but passes through a thin foil. The blue line indicates the final stopping point of the particle in a semi-infinite material, while the red line indicates the border of a 50 µm moderator. The color scale is time since exiting the RF cavity.}
    \label{fig:BackscatteringExample}
\end{figure}

In a thin foil, the positrons do not follow the Makhovian distribution exactly. As an example, Fig.~\ref{fig:BackscatteringExample} shows the trace of a particle with 900 keV kinetic energy. This particle would eventually come to rest at 38 µm (blue line) and contributes to the Makhovian distribution at that point. However, in a 50 µm foil (red line), it would escape before it could scatter backwards and stop.

\begin{figure}[htbp!]
    \centering
    \includegraphics[width=\linewidth]{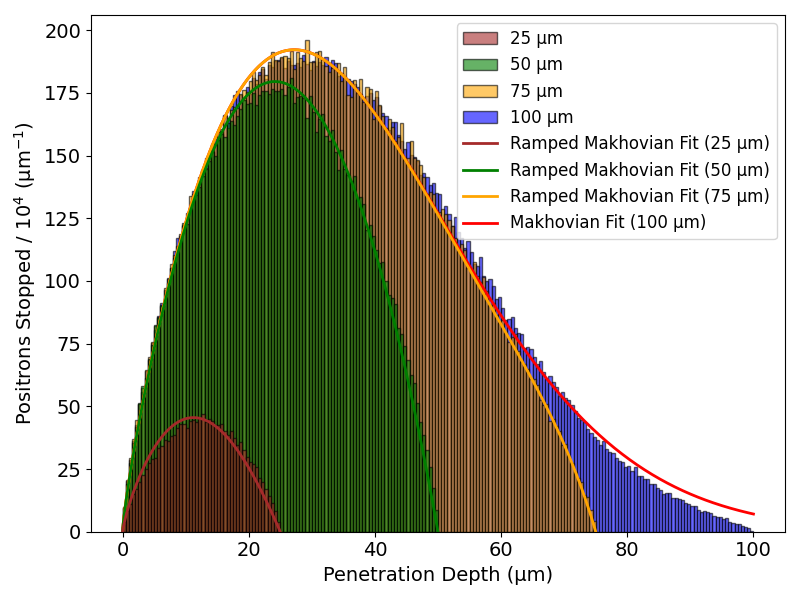}
    \caption{Simulation (shaded) and ramped Makhovian fits for 500 keV positrons on foils of various thicknesses. The red line is an unramped Makhovian for the 100 µm foil.}
    \label{fig:RampedMakhovian}
\end{figure}

If the probability of a particle escaping increases exponentially to 100\% at the border, the Makhovian distribution should be attenuated by an exponential ramp
\begin{equation}
    \tilde{p}(z)=p(z)\left( 1-\exp\left( -\frac{L-z}{\lambda} \right) \right),
\end{equation}
where $L$ is the moderator length and $\lambda$ is a constant dependent on $L$. This function is a good approximation of the distribution for thin foils, as shown in Fig.~\ref{fig:RampedMakhovian}. The fit parameters for Fig.~\ref{fig:RampedMakhovian} are listed in Table~\ref{tab:RampedMakhovian}.

\begin{table}
\centering
\caption{Ramped Makhovian distribution parameter $\lambda$ for 500 keV positrons on a thin tungsten foil}
\label{tab:RampedMakhovian}
\begin{ruledtabular}
\begin{tabular}{cccc}
&Foil Thickness (µm) & $\lambda$ (µm) \\
\hline
&25 & $32.94\pm0.42$ & \\
&50 & $9.00\pm0.12$ & \\
&75 & $4.65\pm0.21$ &  \\
\end{tabular}
\end{ruledtabular}
\end{table}

\subsection{Positron Thermalization}
Due to the negative work function of tungsten, the positrons that stop close to the surface of the foil can be ejected from the moderator. The number of ejected positrons is characterized by a diffusion length $L_+$. For polycrystalline tungsten, $L_+=55$ nm. Single-crystal tungsten has a higher diffusion length ($L_+=135$ nm) but cannot be manufactured in a thin foil, posing a challenge for traditional moderator designs.

The probability of thermalization for a particle is given by
\begin{equation}
    P_\text{thermalization}(z)=\exp\left( -\frac{z}{L_+} \right),
\end{equation}
with $z$ representing the distance to the surface~\cite{ORourke}.

This can also be written as an integral over the stopping distribution $\tilde{p}(z)$,
\begin{equation}
    \label{eq:therm}
    P_\text{thermalization}=\int_0^L \tilde{p}(z)\left[ e^{-\frac{z}{L_+}}+e^{-\frac{L-z}{L_+}} \right]dz,
\end{equation}
where $L$ is the length of the moderator.

As GEANT4 is unable to adequately model positron thermalization, we measure the stopping distribution $p(z)$ by applying a 50 eV kinetic energy cutoff and use Eq.~\ref{eq:therm} to determine the moderator yield.

\section{Moderator Designs}
\subsection{Traditional Foil Moderator}
The first moderator design tested was a single 50 µm polycrystalline tungsten foil with no AMD or RF cavity.

\begin{figure}[htbp!]
    \centering
    \subfloat[Before RF cavity\label{fig:FoilStop2}]
        {\includegraphics[width=0.48\linewidth]{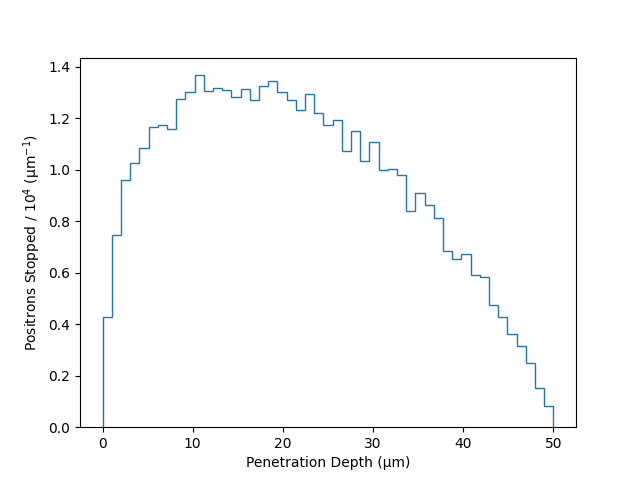}}\hfill
    \subfloat[After RF cavity\label{fig:FoilStop3}]
        {\includegraphics[width=0.48\linewidth]{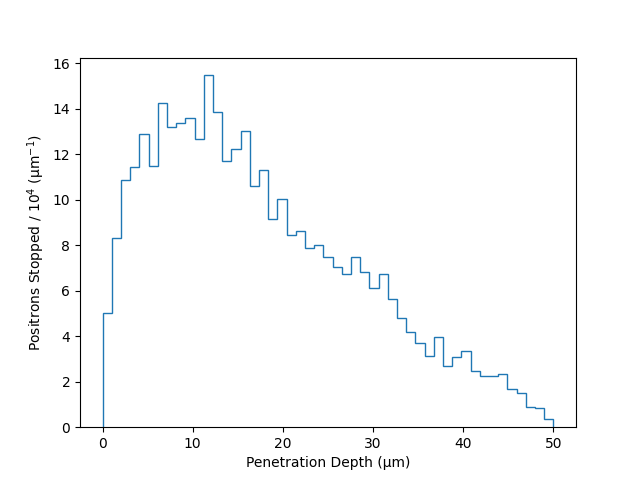}}
    \caption{Stopping distributions of positrons at normal incidence, using the energy values from Fig.~\ref{fig:energytarget}.}
    \label{fig:FoilStopCombined}
\end{figure}

Fig.~\ref{fig:FoilStop2} shows the stopping distribution the full energy spread generated in Fig.~\ref{fig:energytarget}. From Eq.~\ref{eq:therm}, we can calculate the expected positron yield $\gamma_m$,
\begin{equation}
    \gamma_m=\frac{1}{N}\sum_N P_\text{thermalization},
    \label{eq:posyield}
\end{equation}
where $N$ is the number of simulated particles. For the 50 µm foil, simulations show that $\gamma_m=1.49\pm0.55\times10^{-7}$. This represents the fraction of positrons produced by the target that thermalize and are reemitted by the moderator.

The overall efficiency of electrons converted to positrons was $3.81\pm1.41\times10^{-7}\text{ e}^+/\text{e}^-$. This compares to approximately $16\times10^{-7}\text{ e}^+/\text{e}^-$ for a positron source at LLNL~\cite{LLNLpositron} and $0.28\times10^{-7}\text{ e}^+/\text{e}^-$ for the GBAR experiment,~\cite{GBAR} though both setups differed from the design simulated here.

\subsection{Foil with AMD and RF Cavity}

\begin{figure}[htbp!]
    \centering
    \includegraphics[width=\linewidth]{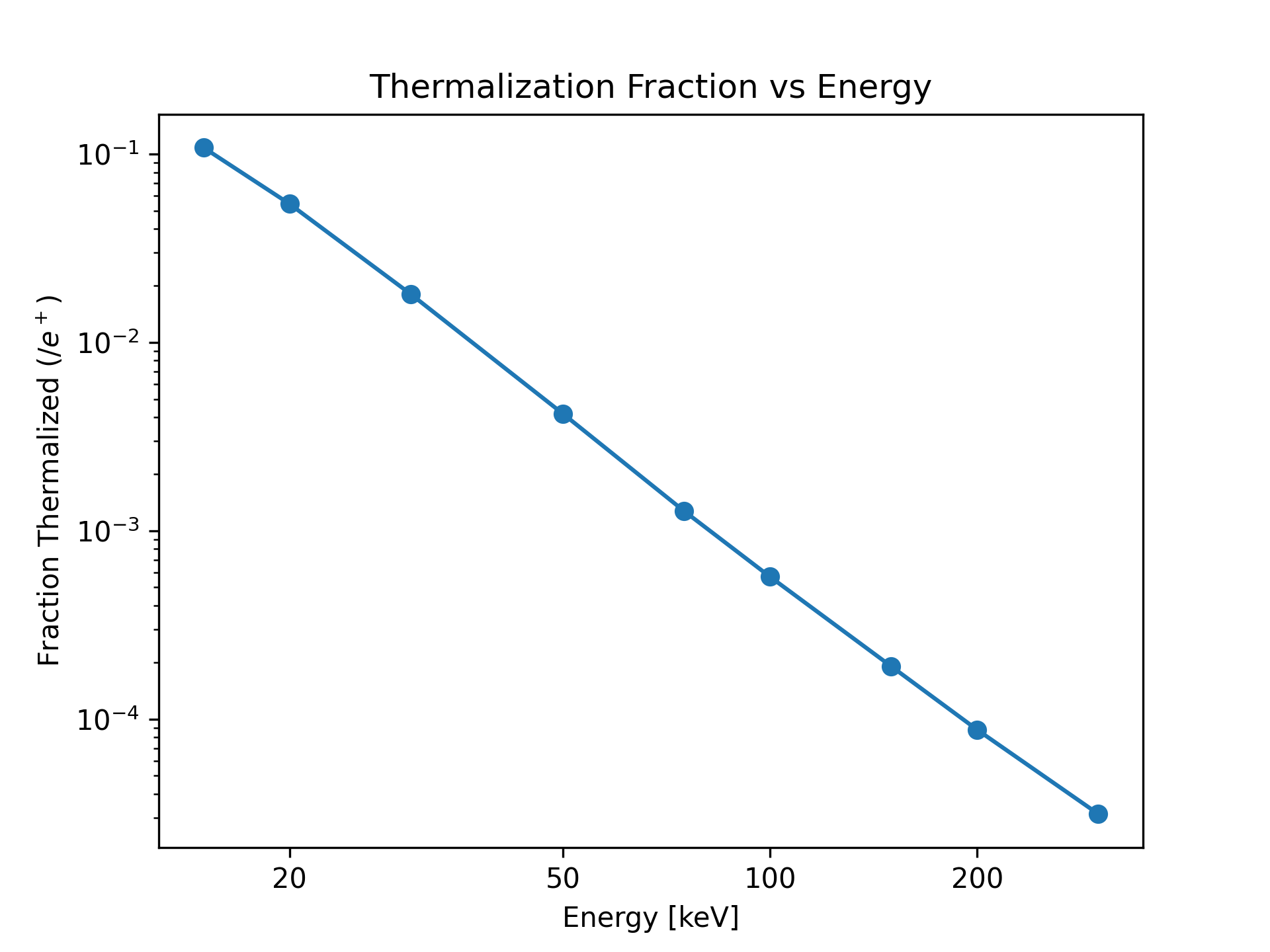}
    \caption{Fraction of positrons reemitted from a 50 µm tungsten foil moderator as a function of energy. Lower energy positrons are significantly more likely to be reemitted. This plot uses the theoretical Makhovian distributions from Fig.~\ref{fig:MakhovianParams}.}
    \label{fig:ThermFrac}
\end{figure}

High-energy positrons do not stop in the moderator, as shown in Fig.~\ref{fig:FoilEnergy} and Fig.~\ref{fig:ThermFrac}. Therefore, the decelerating RF cavity can increase the yield by lowering the beam energy. The new stopping distribution (Fig.~\ref{fig:FoilStop3}) has a yield of $\gamma_m=3.76\pm0.35\times10^{-6}$. This is approximately a 25.3$\times$ improvement over the case without the AMD and cavity. However, some positrons are lost during the intermediate stages, so the system efficiency is $6.2\pm0.57\times10^{-6}\text{ e}^+/\text{e}^-$, a 16.3-fold improvement.

\subsection{Grid Moderator}
\begin{figure}[htbp!]
    \centering
    \includegraphics[width=\linewidth]{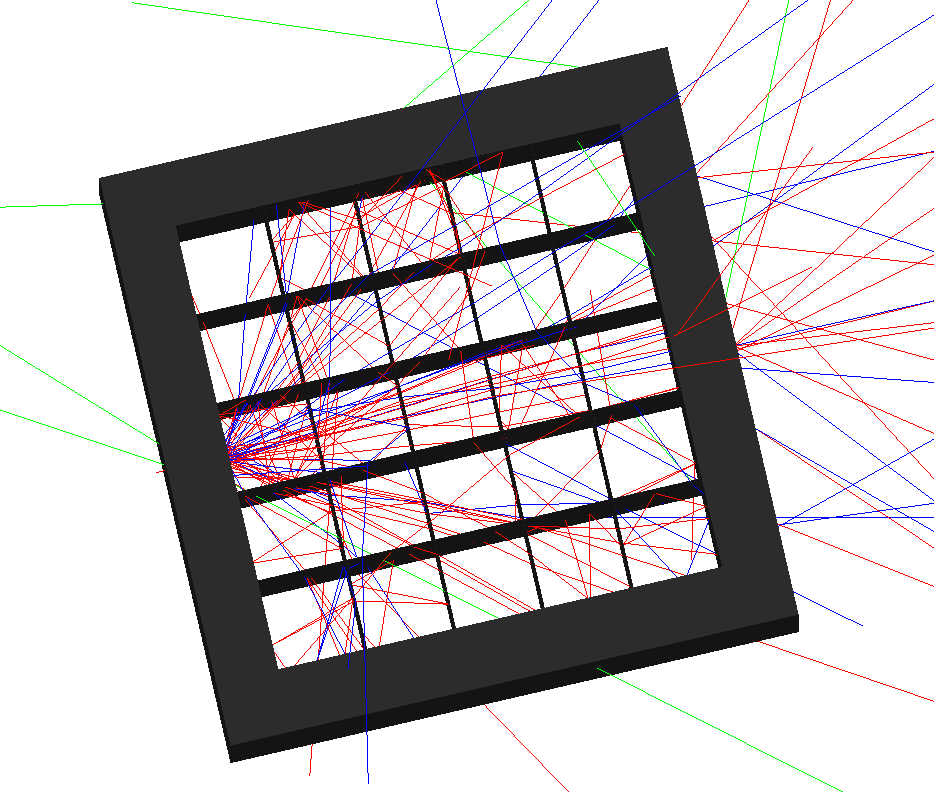}
    \caption{Render of the moderator modeled off of the Slow Positron Facility at KEK. Electrons (red) hit the target from the leftmost edge and the resulting positrons (blue) stop in the mesh. Green lines represent photons.}
    \label{fig:GridModerator}
\end{figure}

In addition to the foil, we simulate a mesh of foils based on of the moderator used at the Slow Positron Facility at KEK.~\cite{KEK} We use four planes of 25 µm foils along each axis (separated 10~mm gaps), surrounded by 7 mm of tungsten to act as the target (Fig.~\ref{fig:GridModerator}). This moderator had a yield of $\gamma_m=3.32\pm 0.26\times10^{-5}$, more than twenty times greater than the foil moderator. However, the grid moderator comes at a cost of increased beam size, since the positrons are spread out over the entire moderator volume.

\subsection{Single-Crystal Moderator}
The final design tested was a single-crystal cylinder of tungsten. This has the advantage of a greater diffusion length, so a greater fraction of positrons at the surface is reemitted. However, this material cannot be a thin foil and is instead a block, meaning positrons can only emerge from one side of the moderator. The efficiency of this moderator was $\gamma_m=3.46\pm0.84\times10^{-6}$.

When simulating the stopping distribution, the single-crystal moderator used the same material as the polycrystalline tungsten since GEANT4 only provides one material class. Instead, the thermalization probabilities were modified.

\section{Discussion}
\begin{figure}[t!]
    \centering
    \includegraphics[width=\linewidth]{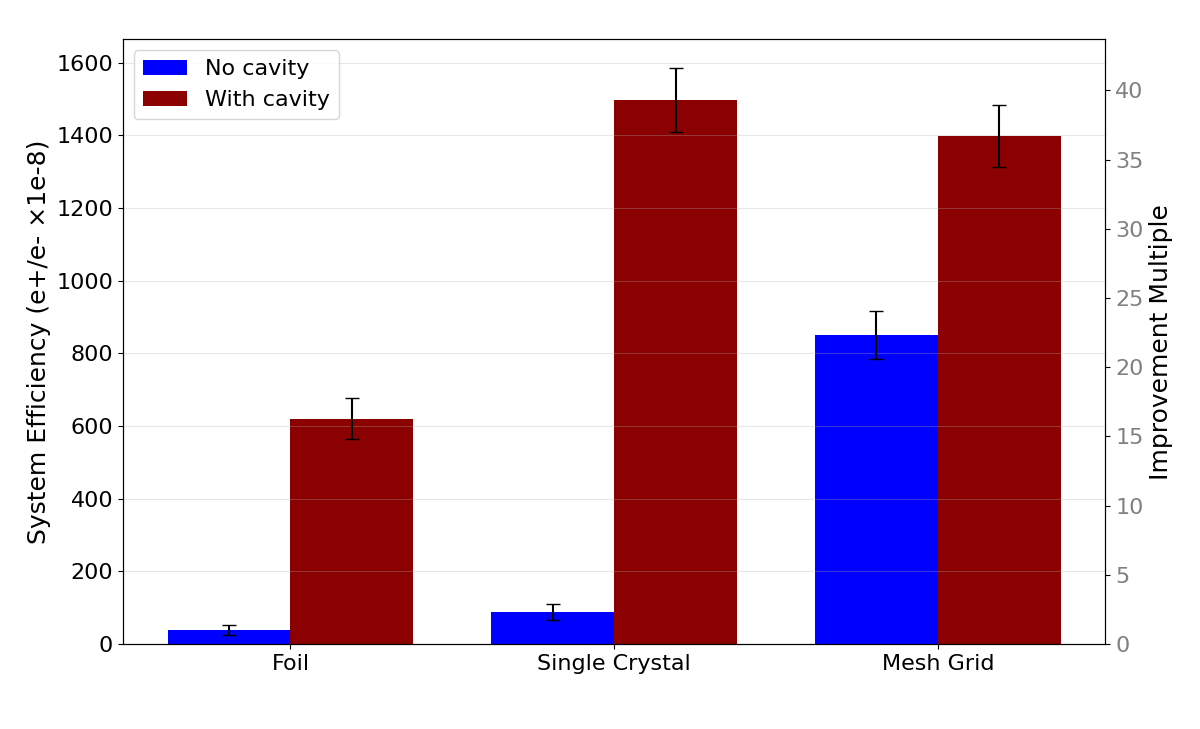}
    \caption{System efficiency of various moderator geometries, with and without the AMD and RF cavity. The right axis shows the improvement multiple in comparison to the foil moderator without a cavity.}
    \label{fig:Summary}
\end{figure}

The overall improvement to system efficiency is shown in Fig.~\ref{fig:Summary}. Adding the linac significantly improves efficiency to all geometry designs. In addition, we find that the single-crystal moderator is the best design for maximizing the yield of the positron source. With the linac, it has a 39.3$\times$ improvement over the traditional method of a simple polycrystalline foil moderator.

\section{Acknowledgements}
This work was supported by the Department of Energy, Laboratory Directed Research and Development program at SLAC National Accelerator Laboratory, under contract DE-AC02-76SF00515. This work was supported in part by the U.S. Department of Energy, Office of Science, Office of Workforce Development for Teachers and Scientists (WDTS) under the Science Undergraduate Laboratory Internships Program (SULI).

\FloatBarrier
\balance

\bibliographystyle{aip}
\bibliography{paper}

\end{document}